\documentclass[prl,twocolumn,showpacs,twoside,floatfix]{revtex4}
\usepackage{graphicx,amsmath,amssymb,mathptm}
\begin{document}
\title{On the evolution of scale-free graphs}
\author{D.-S.~Lee, K.-I.~Goh, B.~Kahng, and D.~Kim}
\affiliation{
\mbox{School of Physics and Center for Theoretical Physics,
Seoul National University, Seoul 151-747, Korea}
}
\date{\today}
\begin{abstract}
We study the evolution of random graphs where edges are added one by one 
between pairs of weighted vertices 
so that resulting graphs are scale-free with the degree exponent $\gamma$. 
We use the branching process approach to 
obtain scaling forms for the cluster size distribution and the 
largest cluster size as functions of the number of edges $L$ and 
vertices $N$. We find that the process of forming a spanning cluster is 
qualitatively different between the cases of $\gamma>3$ and $2<\gamma<3$. 
While for the former, a spanning cluster forms abruptly at a 
critical number of edges $L_c$, generating a single 
peak in the mean cluster size $\langle s \rangle$ as a function 
of $L$, for the latter, however, the formation of a spanning cluster 
occurs in a broad range of $L$, generating double peaks in
$\langle s \rangle$. 
\end{abstract}

\maketitle
Recently, many studies have been performed on
complex networks. Such studies are mostly influenced by the random
graph theory proposed by Erd\H{o}s and R\'enyi (ER)~\cite{er61}. In
the ER model, $N$ number of vertices are present from the
beginning and edges are added one by one in the system, connecting 
pairs of vertices selected randomly.
A remarkable result ER obtained is that 
a giant cluster of size ${\cal O}(N)$, a spanning cluster, 
appears abruptly when $L$ reaches its threshold
value $L_c$, which is  ${\cal O}(N)$. 
Note that the formation of such a spanning cluster can be viewed as a
percolation transition occurring at the
critical probability $p_c=2L_c/N(N-1)={\cal O}(1/N)$.
While the ER graph is pioneering, it is too random, and various properties
of the ER graph are not in accordance with those of complex
networks recently discovered in real world. For example, the
distribution of the number of edges incident on each vertex,
called the degree distribution, is Poissonian for the ER graph,
while it follows a power law for many real-world networks,  
called scale-free (SF) networks~\cite{ba99,mendes02,newman03}.

It was shown that a SF network can be generated by following a
similar way to used in the ER model~\cite{goh01,caldarelli02}. 
$N$ number of vertices are
present from the beginning and edges are added one by one. For
SF networks, however, each vertex with the index of an integer $i$
($i=1,\cdots, N$) is not identical, but is assigned a normalized
weight $w_i=i^{-\alpha}/\sum_{j=1}^N j^{-\alpha}$ with a control parameter
$\alpha \in [0,1)$. Each edge connects a pair of vertices $(i,j)$
selected with probability $w_i w_j$. Thus the ER graph is 
generated with $\alpha=0$. The process of adding edges is
repeated until the total number of edges in the system reaches
$L$. 
This process of constructing networks is called the static model.
When $L$ is in the intermediate regime 
$L_{\ell} \ll L \ll L_{u}=L_{\ell} N$, 
with $L_{\ell}$ being specified below,  
the degree distribution follows a
power law, $p_d(k) \sim k^{-\gamma}$ with $\gamma=1 + 1/\alpha\in (2,\infty)$. 
However, when $L\lesssim L_{\ell}$ ($L\gtrsim L_{u}$),
the network is too sparse (dense), so that the degree distribution
does not follow the SF behavior. 
The static model was introduced to generate SF networks with
various $\gamma$, being used to study various problems.
However, it has not been studied yet how clusters
evolve as the number of edges $L$ increases, which is 
the goal of this Letter.

\begin{figure}
\includegraphics[width=8.0cm]{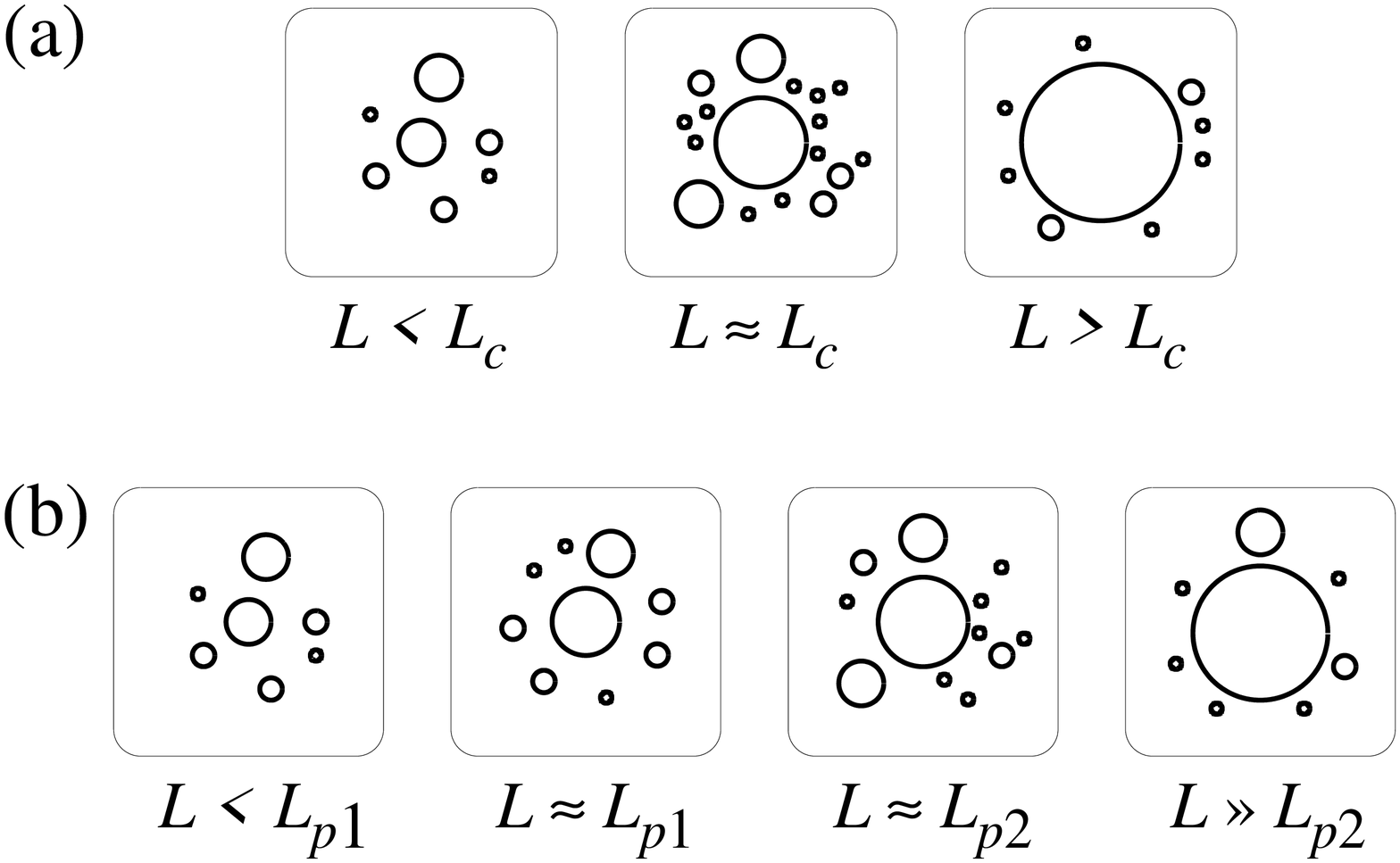}
\caption{Schematic picture for the comparison of cluster evolution
between (I,II)~(a) and (III)~(b).}
\label{fig:evolution}
\end{figure}

\begin{figure}
\includegraphics[width=8.0cm]{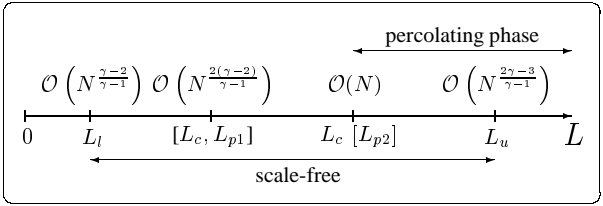}
\caption{Schematic phase diagram of the static model. The
SF behavior of the degree distribution appears between
$L_{\ell}$ and $L_u$. A spanning cluster emerges at $L_c$ for
(I,II), and around $L_{p2}$ for (III). 
The quantities in [...] are only for (III). } 
\label{fig:phase}
\end{figure}

The percolation problem of SF networks has been 
studied~\cite{albert00,cohen00,callaway00,cohen02}, 
reversely, that is, by removing randomly-selected vertices as well 
as their attached edges. 
In this Letter, we study how the cluster evolution of SF graphs 
proceeds as edges are added. Besides confirming the previous 
results in Ref.~\cite{cohen00,cohen02},  
we show that the process of forming a spanning cluster for the 
case of $2<\gamma<3$ is fundamentally different from that of $\gamma>3$.
When $\gamma>3$, as in the case of the ER graph, there exists a critical 
number of edges $L_c$ at which a spanning cluster forms through many 
small clusters coalescing, and the mean cluster size diverges at finite 
$L_c/N$ in the thermodynamic limit. 
In other words, a percolation transition occurs at $L_c$.
When $2 < \gamma < 3$, however, large or small clusters grow in a similar 
manner as a whole without  sudden coalescence occurring. 
As a result, the mean cluster size does not diverge anywhere, but instead  
exhibits two peaks at $L_{p1}$ and $L_{p2}$. Near $L_{p1}$, some small 
clusters merge together forming a much larger one, but it does not span 
the entire system. After passing $L_{p1}$, the largest cluster as well as 
smaller ones continue to grow, and the largest one becomes 
as large as ${\cal O}(N)$ around $L_{p2}$. Throughout this Letter, 
we will denote the case of $\gamma > 4$ as (I), $ 3 < \gamma < 4$ as (II), 
and $ 2 < \gamma < 3$ as (III). The schematic diagram of the 
cluster formation is shown in Fig.~\ref{fig:evolution}.
We obtain characteristic numbers of edges for each case as 
a function of $N$ and summarize them in the phase diagram
shown in Fig.~\ref{fig:phase}. Moreover, we derive scaling forms for 
the cluster size distribution and the largest cluster size 
analytically and numerically. 

{\it Power-law degree distribution} --- The probability $p_{d,i}(k)$ that a
vertex $i$ has degree $k$ follows approximately a Poissonian form as 
$p_{d,i}(k)\simeq \langle k_i \rangle^k \exp(-\langle k_i \rangle) /k!$ 
for large $k$ and large $N$.  
Here the average degree of the vertex $i$ is given by 
$\langle k_i \rangle=K(L) i^{-1/(\gamma-1)}$, where 
$K(L)=\langle k_1 \rangle = 2L/\zeta_N[1/(\gamma-1)]$,  
with $\zeta_N (x) \equiv \sum_{j=1}^N j^{-x}$. 
Note that $\zeta_N(x)$ converges to the Riemann zeta function $\zeta(x)$ 
for $x>1$ but scales as $N^{1-x}/(1-x)$ for $x<1$. 
When $L$ is small enough, most vertices
have no edge.
When $\langle k_1 \rangle\sim 1$, that is,
$L \sim L_{\ell}\equiv \zeta_N[1/(\gamma-1)]\sim N^{(\gamma-2)/(\gamma-1)}$, 
the SF behavior in the degree distribution begins to appear.
When $L\gg L_{\ell}$, the degree
distribution is derived as
\begin{equation}
p_d(k)={1 \over N}\sum_{i=1}^N p_{d,i}(k) \simeq  
c\, k^{-\gamma}, 
\label{eq:degree}
\end{equation}
where $c$ is given as $c\simeq (\gamma-1) [K(L)]^{\gamma-1}/N$.


{\it Branching process approach} --- As $L$ increases beyond $L_{\ell}$, 
small clusters form.
However, clusters are still sparse and
of tree structure~\cite{er61}. 
The formation of such sparse clusters 
can be understood through the
multiplicative branching process approach~\cite{harris89}.
Here we introduce the probability distribution $P(s)$  as 
the number of vertices  
belonging to $s$-clusters, clusters with $s$ vertices, 
divided by $N$~\cite{aharony92}. 
Also we define another probability distribution $R(s)$ as 
the number of edges followed by  
 $s$-clusters  divided by $2L$.
The generating functions of those quantities
are defined as ${\cal P}(z) =\sum_s P(s) z^s$ and ${\cal
R}(z) =\sum_s R(s) z^s$, respectively.
Both summations run over finite
clusters only~\cite{harris89}, and then when clusters are sparse 
the following relations hold:  
\begin{equation}
{\cal R}(z)=z f({\cal R}(z)) \quad  {\rm and}
\quad {\cal P}(z)=z g({\cal R}(z)),
\label{eq:sc}
\end{equation}
where $g(\omega)=\sum_{k=0}^\infty p_d(k)\omega^k$ and 
$f(\omega)=g'(\omega)/\langle k \rangle$ with 
$\langle k^n \rangle=\sum_{k=0}^\infty k^n p_d(k)$. 

To apply Eq.~(\ref{eq:sc}) to the static model, we 
use the following form valid in the limit $1-\omega\ll 1$; 
$g(\omega)\simeq(1/N)\sum_{i=1}^N \exp[\langle k_i \rangle(\omega-1)]$.
Then it is obtained that  
$z=\omega+\sum_{n=1}^\infty a_n (1-\omega)^n$
for $K(L)(1-\omega)\leq 1$, where $a_1=f'(1)$, $a_2$ a negative constant, 
and so on,
while 
\begin{equation}
z=\omega +\sum_{n=1}^{\lfloor \gamma-2\rfloor}
a_n(1-\omega)^n
+ A (1-\omega)^{\gamma-2} + \ldots
\label{eq:z2}
\end{equation}
for $K(L)(1-\omega)\gg 1$, 
where $\lfloor x \rfloor $ is the floor
function of $x$ and $A=\Gamma(3-\gamma) 
[K(L)/N^{1/(\gamma-1)}]^{\gamma-2}$. 
The generating function $\omega={\cal R}(z)$ can be obtained by
inverting $z=\omega/f(\omega)$, and ${\cal P}(z)$ is then obtained
by using ${\cal P}(z)=z g ({\cal R}(z))$.

{\it Critical point} --- The values of ${\cal P}(1)$ and ${\cal R}(1)$ 
are $1$ only when $f'(1)\leq 1$, while they are smaller than $1$ 
when $f'(1)>1$. Thus the condition $f'(1)=1$, which is the same as 
the condition $\langle k^2 \rangle /\langle k \rangle=2$~\cite{molloy95}, 
leads to a characteristic number of edges $L_c$,
\begin{equation}
L_c ={1 \over 2} {\zeta_N[1/(\gamma-1)]^2 \over \zeta_N[2/(\gamma-1)]}.
\label{eq:lc}
\end{equation}
For large $N$, $L_c$ is $N\, (\gamma-1)(\gamma-3)/(2(\gamma-2)^2)$ 
for (I,II), 
and $N^{2(\gamma-2)/(\gamma-1)}(\gamma-1)^2/(2(\gamma-2)^2\zeta_N
[2/(\gamma-1)])$ for (III). 
We will show later that  for (I,II), a spanning cluster appears at $L_c$, 
but for (III), the size of the largest cluster does not reach 
${\cal O}(N)$ at $L_c$. Thus more edges are needed to generate 
a spanning cluster. 

{\it Cluster size distribution} --- The asymptotic behaviors of 
$P(s)$ and $R(s)$ for large $s$ can be obtained from the singular parts of
${\cal P}(z)$ and ${\cal R}(z)$ as $z \rightarrow 1$,
respectively. In the static model, 
the characteristic behaviors of ${\cal P}(z)$ and ${\cal R}(z)$
depend on the degree exponent $\gamma$, classifying them into the three cases, 
(I), (II), and (III).
Each case is again
classified into  (i) subcritical ($L < L_c$), (ii)
critical ($L=L_c$), and (iii) supercritical cases ($L >
L_c$). Our results for $P(s)$ are listed in Table I for each case.
\begin{table}
\caption{
Cluster size distribution $P(s)$.
Here $\Delta \equiv (L-L_c)/L_c$, and 
the scaling exponents $(\tau,\sigma)$  are given as $(3/2,1/2)$
for (I) and $({(\gamma-1)/(\gamma-2)},{(\gamma-3)/(\gamma-2)})$
for (II).}
\label{table:summary}
\begin{ruledtabular}
\begin{tabular}{c|c|c|c}
&(i)&(ii)&(iii)\\
&$(L<L_c)$&$(L=L_c)$&$(L>L_c)$\\
\hline
(I,II)&
$
\begin{array}{lr} s^{-\tau}& (s\ll s_c)\\
(|\Delta|s)^{-(\gamma-1)}&(s\gg s_c)\\
\end{array}
$& 
$s^{-\tau}$&
$
\begin{array}{lr}
s^{-\tau}& (s\ll s_c)\\
\exp(-s/s_c)&(s\gg s_c)\\
\end{array}
$ 
\\ 
\cline{2-4}
&\multicolumn{3}{c}{
$s_c\sim |\Delta|^{-1/\sigma}$}\\
\hline
\hline
(III)&\multicolumn{2}{c|}{
$
\begin{array}{lr}
(Ns/L)^{-(\gamma-1)}& (s\ll K(L))\\
N^{-{4-\gamma\over 2(\gamma-1)}}
e^{-s/s_c}& (s\gg K(L))
\end{array}
$}&
$
\begin{array}{lr}
(Ns/L)^{-(\gamma-1)}& (s\ll s_c)\\
\left({L/N}\right)^{4-\gamma\over 2(3-\gamma)} 
e^{-s/s_c}& (s\gg s_c)
\end{array}
$ \\ 
\cline{2-4}
&\multicolumn{2}{c|}{
$s_c\sim L^2/(N^{3(\gamma-2)/(\gamma-1)}|\Delta|^2)$}&
$s_c\sim (L/N)^{-(\gamma-2)/(3-\gamma)}$
\end{tabular}
\end{ruledtabular}
\end{table}

{\it Emergence of spanning cluster} --- 
In cases of (i) and (ii),
the size of the
largest cluster $S$ is obtained self-consistently through the relation, 
$\sum_{s \ne S} P(s)=1-{S/N}$ 
using $P(s)$ in Table~\ref{table:summary}. 
For example, when
$P(s)\sim s^{-\tau}$, $S$ is obtained to be $S\sim N^{1/\tau}$.
Thus when $L=L_c$ (ii),  
$S\sim N^{2/3}$ for (I)~\cite{er61} and $\sim
N^{(\gamma-2)/(\gamma-1)}$ for (II)~\cite{cohen_book}.  
For (III),
using $P(s)\sim (Ns/L_c)^{1-\gamma}$, we obtain 
$S\sim N^{(\gamma-2)/(\gamma-1)}$, but 
we show below that this is not the incipient 
spanning cluster.
The size of the largest cluster for the subcritical case (i) 
is 
$S\sim \max\{K(L)/|\Delta|,|\Delta|^{-1/\sigma}\}$ for (I), 
$\sim K(L)/|\Delta|$ for (II), 
and $\sim K(L)$ for (III), 
with $\Delta=(L-L_c)/L_c$. 

In case of (iii),  the theory of the multiplicative branching 
process yields the size of 
an {\it infinite} cluster  
through $N(1-{\cal P}(1))$.  
Thus it can be identified 
with the largest cluster if it is 
larger than $S$ at $L_c$.  
From Eq.~(\ref{eq:sc}), 
$1-{\cal P}(1)\sim (2L/N)(1-{\cal R}(1))$,  
and the value of $1-{\cal R}(1)=1-\omega$ is obtained 
by solving Eq.~(\ref{eq:z2}) with $z=1$. 
Thus we obtain, $S\sim \Delta^\beta N$ with $\beta=1$ for (I) and 
$\beta=1/(\gamma-3)$ for (II) in the regime of $\Delta N^{1/\mu}\gg 1$,  
while 
$S\sim (L/N)^{1/(3-\gamma)} N$  for (III)  when $\Delta\gg 1$,  
where a new scaling exponent $\mu=3$ for (I) and $(\gamma-1)/(\gamma-3)$
for (II) was used. The behaviors of (i), (ii), and (iii) lead 
to a scaling ansatz for (I,II),
\begin{equation}
S \sim N^{1/\tau}~\Psi_{\rm (I,II)} (\Delta N^{1/\mu}),
\label{eq:S1}
\end{equation}
where $\tau$ is given  in Table~\ref{table:summary},  and 
$\Psi_{\rm (I,II)} (x)$ is constant for $|x| \ll 1$ and behaves as
$x^{\beta}$ for $x \gg 1$ and $|x|^{-\delta}$ for $x\ll -1$ with 
$\delta=2$ for (I) and $1$ for (II). 
In the thermodynamic limit, a spanning cluster 
emerges if only $L\geq L_c$. On the other hand, for (III), 
\begin{equation}
S\sim N^{(\gamma-2)/(\gamma-1)}~ \Psi_{\rm (III)}(\Delta),
\label{eq:S2}
\end{equation}
where $\Psi_{\rm (III)}(x)$ is a constant for $|x|\lesssim 1$, behaving as 
$(1+x)$ when $x\simeq -1$ and $x^{1/(3-\gamma)}$ for $x\gg 1$. 
The size of the largest cluster is ${\cal O}(N^{(\gamma-2)/(\gamma-1)})$, 
which is not as large as ${\cal O}(N)$, even when $L>L_c$,  so that 
$L_c={\cal O}(N^{2(\gamma-2)/(\gamma-1)})$ is not 
a percolation threshold. 
The largest cluster size $S$ becomes ${\cal O}(N)$ only when $L$ 
becomes as large as ${\cal O}(N)$.

\begin{figure}
\includegraphics[width=8.0cm]{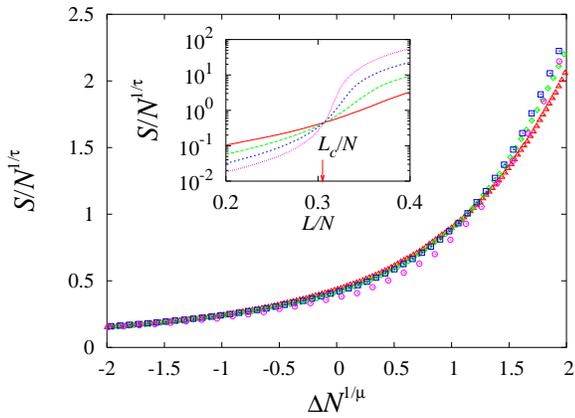}
\caption{Data collapse of  $S/N^{1/\tau}$
versus $\Delta N^{1/\mu}$ for 
Eq.~(\ref{eq:S1}) 
with $\gamma=3.6$, 
and $N=10^4(\triangle)$, $10^5 (\Diamond)$, $10^6 (\square)$, and 
$10^7 (\bigcirc)$.
Here $\tau=(\gamma-1)/(\gamma-2)=13/8$, $\mu=(\gamma-1)/(\gamma-3)=13/3$, 
and $L_c/N\simeq 0.305$ 
from Eq.~(\ref{eq:lc}) are used. 
Inset: the same data plotted versus $L/N$ cross at $L/N\simeq L_c/N$. 
}
\label{fig:S1}
\end{figure}

\begin{figure}
\includegraphics[width=8.0cm]{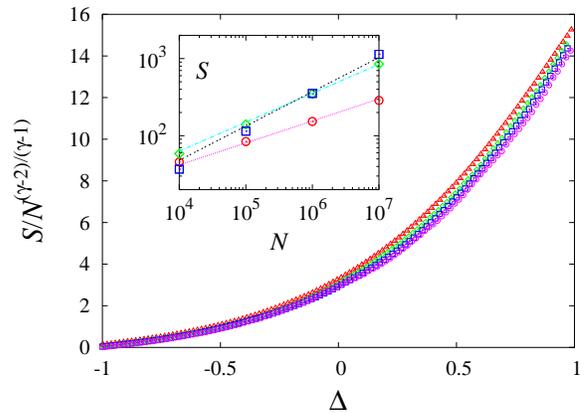}
\caption{Data collapse of $S/N^{(\gamma-2)/(\gamma-1)}$ versus $\Delta$ 
for Eq.~(\ref{eq:S2}) 
with $\gamma=2.4$  and 
$N=10^4(\triangle)$, $10^5 (\Diamond)$, $10^6 (\square)$, and 
$10^7 (\bigcirc)$. Here $L_c/N^{4/7}\simeq 2.08$ from 
Eq.~(\ref{eq:lc}) is used. 
Inset: plot of $S$ at $L_c$  versus $N$ for 
$\gamma=2.4 (\circ)$, $2.6 (\Diamond)$, and $2.8 (\square)$, being 
in accordance with 
$S\sim N^{(\gamma-2)/(\gamma-1)}$ represented by the 
dotted, dashed-dotted, and dashed line, respectively. 
}  
\label{fig:S2}
\end{figure}

To check such scaling behaviors
of $S$, we perform numerical simulations 
for $\gamma=3.6$ and $\gamma=2.4$.
As shown in Figs.~\ref{fig:S1} and ~\ref{fig:S2}, 
the data of $S/N^{(\gamma-2)/(\gamma-1)}$ 
with different $N$
collapse 
with the scaling variables, 
$\Delta N^{1/\mu}$ 
for $\gamma=3.6$ (II), 
and $\Delta$ for $\gamma=2.4$ (III), respectively.   
For $\gamma=3.6$, $L_c/N\simeq 0.305$ 
theoretically obtained in Eq.~(\ref{eq:lc}) 
is confirmed by the data crossing at $L_c/N\simeq 0.306(2)$. 

\begin{figure}
\includegraphics[width=8.0cm]{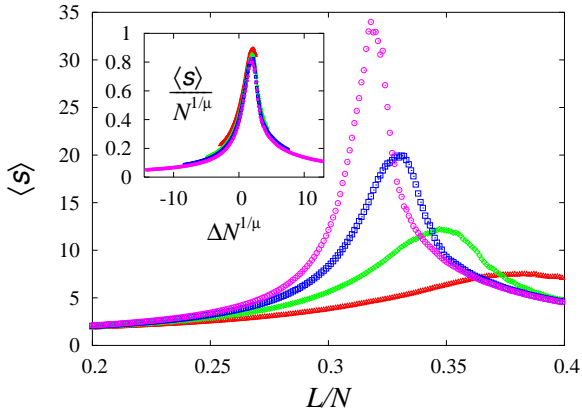}
\caption{Mean cluster size  $\langle s
\rangle$  as a function of $L/N$  
with $\gamma=3.6$ 
for $N=10^4(\triangle)$, $10^5 (\Diamond)$, $10^6 (\square)$,  and 
$10^7 (\bigcirc)$.
The peak heights increase with $N$.   
The inset shows the data collapse of    
the rescaled mean cluster size 
$\langle s \rangle/N^{1/\mu}$ versus 
$\Delta N^{1/\mu}$.} 
\label{fig:sav}
\end{figure}

{\it Mean cluster size} --- The difference in the cluster
evolution for (I), (II),  and (III) appears more apparently
in the mean cluster size $\langle s \rangle$ defined as
$\langle s \rangle \equiv \sum_{s \neq S} sP(s).$
The quantity $\langle s \rangle$ is similar to the susceptibility defined in
the percolation theory but here we exclude the largest cluster 
even for $L < L_c$. For (I,II), as $L$ increases, many small
clusters grow by attaching edges, which continues up to $L=L_c$, 
and then a spanning 
cluster forms by the abrupt coalescence of those small clusters 
as shown in Fig.~\ref{fig:evolution}.
Since we do not count the spanning cluster in calculating $\langle s
\rangle$, $\langle s\rangle$ decreases rapidly as $L$ passes $L_c$. Thus the
mean cluster size exhibits a peak at $L=L_c$, which diverges in
the thermodynamic limit $N\to\infty$.  
The scaling behaviors of $S$ and $P(s)$ in 
Table~\ref{table:summary} lead to another scaling ansatz 
\begin{equation}
\langle s \rangle = N^{1/\mu}~ \Phi (\Delta N^{1/\mu}),
\label{eq:sav}
\end{equation}
where $\Phi(x)$ is a constant when $|x|\ll 1$, and behaves as 
$x^{-1}$ when $x\gg 1$, and $|x|^{-1}$ for (I) and  
$|x|^{-(\gamma-3)/(\gamma-2)}$ for (II)  when $x\ll -1$. 
Such behaviors can be confirmed with numerical data for 
$\gamma=3.6$ in 
Fig.~\ref{fig:sav}. 

\begin{figure}[t]
\includegraphics[width=8.0cm]{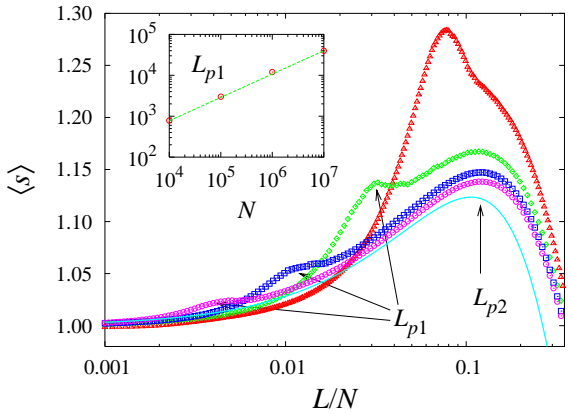}
\caption{Mean cluster size $\langle s \rangle$ 
as a function of $L/N$ in semi-logarithmic scales 
with $\gamma=2.4$ 
for $N=10^4(\triangle)$, $10^5 (\Diamond)$, $10^6 (\square)$,  and 
$10^7 (\bigcirc)$.
In addition to the peak at $L_{p1}$, 
another peak is shown at $L_{p2}\simeq 0.1 \, N$ for  $N=10^5$, $10^6$, and
$10^7$, respectively. 
The measured values of $L_{p1}$ ($\bigcirc$)  
as a function of $N$ are plotted in the inset 
together with the guide line whose slope is $2(\gamma-2)/(\gamma-1)$ 
for comparison. 
The solid line represents Eq.~(\ref{eq:lp2}).} 
\label{fig:double}
\end{figure}

For (III), however, the mean cluster size does not diverge at any
value of $L$ but has two blunt peaks (Fig.~\ref{fig:double}). 
First, it has a small peak
at $L=L_{p1}$,  
but it increases again as $L$
increases beyond $L_{p1}$. Edges newly introduced either create new
clusters of size larger than $1$ or merge small clusters
to the larger one with size not as large as ${\cal O}(N)$. 
When $L$ reaches $L_{p2}={\cal O}(N)$ where  
the second peak arises, 
the largest cluster becomes as large as ${\cal O}(N)$.
When $L$ is near $L_c$, 
$\langle s \rangle$  can be evaluated through 
$\langle s \rangle=\sum_{s\ne S} s P(s)$,  
and it follows that    
$\langle s \rangle-1\sim \min\{s_c^{3-\gamma}, 
S^{3-\gamma}\}$, where $s_c$ is a characteristic cluster size 
defined in Table~\ref{table:summary}, and 
the constant term $1$ originates from 
the isolated vertices whose fraction is nearly $1$. 
Since $S$ ($s_c$) increases (decreases) 
with increasing $L$ for $L>L_c$, $\langle s\rangle$ is maximal 
at $S=s_c$, occurring 
at $L_{p1}=b\,L_c={\cal O}(N^{2(\gamma-2)/(\gamma-1)})$ with 
$b$ being a constant depending on $\gamma$. 
That is verified numerically as shown in the inset of 
Fig.~\ref{fig:double}. For $L_{p1}\ll L \ll N$, 
using $\langle s \rangle = {\cal P}'(1)$, 
we obtain $\langle s \rangle$ to be 
\begin{equation}
\langle s \rangle\simeq 1+{2\over 3-\gamma}({L\over N})-
B\,\left({L\over N}\right)^{1\over 3-\gamma}+
C\, \left({L\over N}\right)^{\gamma-1 \over 3-\gamma},
\label{eq:lp2}
\end{equation}
where $B=(5-\gamma)/(3-\gamma)[2   
((\gamma-2)/(\gamma-1))^{\gamma-2}\Gamma(3-\gamma)]^{1/(3-\gamma)}$ 
and $C=(2/(3-\gamma)[(2(\gamma-2)/
(\gamma-1))^{\gamma-2}\Gamma(3-\gamma)]^{2/(3-\gamma)}- 
\Gamma(2-\gamma)[2  
(\gamma-2)/(\gamma-1)\Gamma(3-\gamma)]^{(\gamma-1)/(3-\gamma)}.$
It exhibits a peak at $L_{p2}/N \simeq 0.1$ for $\gamma=2.4$, 
close to the location obtained by numerical simulations 
(Fig.~\ref{fig:double}). 


{\it Dense graph} --- The power-law degree distribution in
Eq.~(\ref{eq:degree}) 
lasts up to $L\sim L_{u}\equiv \zeta_N[1/(\gamma-1)]N={\cal O}(N^{(2\gamma-3)/(\gamma-1)})$, 
around which  the vertex $i=1$ is connected to nearly all vertices  
($\langle k_1 \rangle \sim N$) and 
the degree distribution $p_d(k)$ begins to develop a peak at $k=N-1$. 
As $L$ increases beyond $L_{u}$, 
more vertices have such maximal number of edges, 
and the graph becomes denser.  

{\it Summary} ---
We have studied how clusters of SF graphs are created and evolve as the 
number of edges increases. 
We obtained the cluster size distribution, the largest cluster size, 
and the mean cluster size as functions of the numbers of edges $L$ and 
vertices $N$. Those quantities behave differently  
when $\gamma>3$ and $2<\gamma<3$. For the former, a giant spanning 
cluster forms through a sudden coalescence of small clusters, 
exhibiting a percolation transition, while for the latter, it does gradually, 
and the mean cluster size shows double peaks at distinct numbers of edges, 
$L_{p1}$ and $L_{p2}$. This result implies that the fragmentation process 
of SF graphs under random failures on edges is qualitatively similar to 
(different from) the one under intentional attack when 
$\gamma > 3$ ($2< \gamma < 3$)~\cite{albert00,cohen00}.  
Finally, it is noteworthy that recently Aiello {\it el al.}~\cite{aiello02} 
studied the possibility of forming a spanning cluster for given 
$N$ and $L=(N/2)\zeta(\gamma-1)/\zeta(\gamma)$ as a function 
of $\gamma$, and found that a spanning cluster 
can exist only for $\gamma < \gamma_c\simeq 3.48$. However, 
the way of constructing a SF graph in their model is different 
from ours.\\

The authors would like to thank S. Havlin for helpful comments on 
the manuscript. This work is supported by the KOSEF Grant 
No. R14-2002-059-010000-0 in the ABRL program.

\end{document}